\documentclass[fleqn,twoside,epsfig]{article}
\usepackage{espcrc2}
\usepackage{amssymb}

\usepackage{graphicx}
\usepackage[figuresright]{rotating}

\newcommand{\AmS}{{\protect\the\textfont2
  A\kern-.1667em\lower.5ex\hbox{M}\kern-.125emS}}
\newcommand{\nc}{\newcommand}
\nc{\be}{\begin{equation}}
\nc{\ee}{\end{equation}}
\nc{\bea}{\begin{eqnarray}}
\nc{\eea}{\end{eqnarray}}
\nc{\3}{12^3\times{24}}
\nc{\slh}{\!\!\!/}
\nc{\cmuofp}{C_{\mu}(p)}
\nc{\aofp}{A(p)}
\nc{\bofp}{B(p)}
\nc{\ccmuofp}{{\cal{C}}_{\mu}(p)}
\nc{\caofp}{{\cal{A}}(p)}
\nc{\cbofp}{{\cal{B}}(p)}
\nc{\caofps}{{\cal{A}}^{2}(p)}
\nc{\cbofps}{{\cal{B}}^{2}(p)}
\nc{\acofp}{A^{(c)}(p)}
\nc{\bcofp}{B^{(c)}(p)}
\nc{\aoofp}{A^{(0)}(p)}
\nc{\boofp}{B^{(0)}(p)}
\nc{\trace}{{\rm{Tr}}}
\nc{\comuofp}{C_{\mu}^{(0)}(p)}
\nc{\ccomuofp}{{\cal{C}}_{\mu}^{(0)}(p)}
\nc{\bib}{\bibitem}
\nc{\nm}{\nonumber}
\nc{\mofp}{M(q^{2})}
\nc{\zofp}{Z(q^{2})}
\nc{\mcofp}{M^{(c)}(q^{2})}
\nc{\zcofp}{Z^{(c)}(q^{2})}
\nc{\link}{U_{\mu}(x)}
\nc{\hm}{\hat{\mu}}
\nc{\hn}{\hat{\nu}}
\nc{\dg}{\dagger}
\nc{\calz}{{\cal{Z}}}

\title{\vspace{-1.8cm}
\hfill \rm \null \hfill
 \hbox{\normalsize ADP-02-66/T506} \hskip1.5cm \phantom{?}\\
\vspace{+1.3cm}
Quark propagator in a covariant gauge}

\author{F.D.R. Bonnet,
        D.B. Leinweber,
        A.G. Williams,
        J.M. Zanotti, and\,
        J.B. Zhang.\\
\vspace{+4mm}Dept. of Physics and Mathematical Physics and CSSM, University of Adelaide, Adelaide 5005.}
\begin{document}

\begin{abstract}
Using mean--field improved gauge field configurations,
we compare the results obtained for the quark propagator from Wilson
fermions and Overlap fermions on a $\3$ lattice at a spacing of $a=0.125(2)$ fm.
\vspace{1pc}
\end{abstract}

\maketitle

\section{INTRODUCTION}

In QCD, asymptotic freedom states that at short distances the effective coupling constant vanishes.
In that regime the interaction between quarks and gluons is largely reduced.
The dynamics of these particles can be studied through their propagators in
momentum space. The quark propagator is really a description of how the
quark propagates in the QCD vacuum and is one of the most fundamental building
blocks of QCD. By studying the momentum dependent quark mass function in the infrared region,
the scalar part of the propagator, we can gain some insights into the mechanism of
chiral symmetry breaking. Chiral symmetry is dynamically broken in the QCD vacuum.
This gives rise to mass generation in the infrared.

In the deep infrared region, artifacts associated with the finite size of the
lattice spacing become small. This is the most interesting region as nonperturbative physics
lies here. However, the ultraviolet behaviour of the propagator at large momenta will in general
strongly deviate from the correct continuum behaviour. This behaviour will be action dependent.

In this brief report we compare the quark propagators of Wilson and overlap fermions. Additional details
may be found in Ref.~\cite{bonnet}.

\subsection{The Quark Propagator}

In the continuum, it is possible to study dynamical chiral symmetry breaking
(DCSB) using the renormalized quark Dyson--Schwinger equation (quark--DSE) in Euclidean space.
In this equation, information about the renormalized dressed gluon propagator and dressed quark--gluon vertex
is embodied. The quark propagator has the general form
\bea
S(p)&=&\frac{1}{i\gamma\cdot{p}A(p^2;\zeta^{2})+B(p^2;\zeta^{2})}\nm\\
&=&\frac{Z(p^2;\zeta^{2})}{i\gamma\cdot{p}+M(p^2;\zeta^{2})},
\label{quarkprop}
\eea
where $M=B/A$ and $Z=1/A$. Here, the parameter $\zeta$ represents the renormalization point.
The functions $A(p^2;\zeta^{2})$ and $B(p^2;\zeta^{2})$ carry all the 
effects of vector and scalar quark dressing induced by the quark 
interactions with the gluon field.

Through this simple form we can extract the quark mass function $M(p)$ 
and renormalization function, $Z(p)$.

\section{QUARK PROPAGATOR ON THE LATTICE}

\begin{figure}[tbh]
\resizebox*{\columnwidth}{!}{\rotatebox{90}{\includegraphics{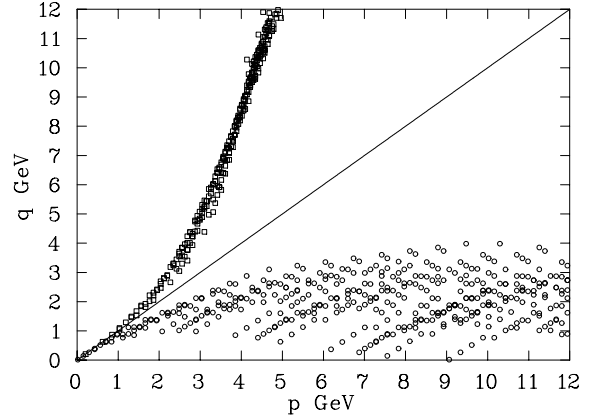}}}
\caption{\label{fig:pvsq}The lattice momentum $q$ versus the discrete momentum $p$, both in GeV. Overlap ($\square$) and Wilson fermion ($\circ$) momenta are indicated.}
\end{figure}

On the lattice we expect the bare quark propagators, in momentum space,
to have a similar form as in the continuum~\cite{jon1,jon2,alkofer}. Hence, the dimensionless inverse lattice
bare quark propagator takes the general form of
\bea
S^{-1}(p)&\equiv&{i\left(\sum_{\mu}\cmuofp\gamma_{\mu}\right)+\bofp}\nm\\
&\equiv&\frac{i\left(\sum_{\mu}\ccmuofp\gamma_{\mu}\right)+\cbofp}{{\cal{C}}^{2}(p)+{\cal{B}}^{2}(p)},
\label{invquargen}
\eea
with ${\cal{C}}^{2}(p)=\sum_\mu(\ccmuofp)^{2}$. The discrete momentum values for a
$t$--antiperiodic lattice of size $N^{3}_{i}\times{N_{t}}$, with $n_i=1,..,N_i$ and $n_t=1,..,N_t$,
are given by $p_i=\frac{2\pi}{N_{i}a}\left(n_i-\frac{N_i}{2}\right)$ and
$p_t=\frac{2\pi}{N_{t}a}\left(n_t-\frac{1}{2}-\frac{N_t}{2}\right)$.
The quark propagator is
\be
S(p)\equiv{-i\left(\sum_{\mu}\ccmuofp\gamma_{\mu}\right)+\cbofp}.
\ee
Taking the trace we obtain
\bea
\ccmuofp&=&\frac{i}{4N_c}{\trace}[\gamma_\mu{S(p)}], \hspace{0.1cm}{\rm{and}}\hspace{0.1cm}\nm\\
\cbofp&=&\frac{1}{4N_c}{\trace}[S(p)],
\eea
which we use to construct the $\cmuofp$ and $\bofp$
\bea
\label{cmup}
\cmuofp&=&\frac{\ccmuofp}{{\cal{D}}(p)},\hspace{0.3cm}{\rm{and}}\hspace{0.3cm}\bofp=\frac{\cbofp}{{\cal{D}}(p)},
\eea
where ${\cal{D}}(p)={{\cal{C}}^{2}(p)+{\cal{B}}^{2}(p)}$. At tree--level the 
lattice quark propagator takes the same form as in Eq.~(\ref{quarkprop}) 
and we know that the free propagator
$(S^{(0)}(p))^{-1}{\equiv}(Z^{(0)}(p))^{-1}\left[ik\slh+M^{(0)}(p)\right]$, 
where $k_{\mu}\longrightarrow{p_{\mu}}$ as $p_{\mu}\longrightarrow{0}$. 
It is then possible
to extract the momentum directly from the lattice by calculating
\bea
q_\mu\equiv\comuofp&=&\frac{\ccomuofp}{({\cal{C}}^{(0)}(p))^{2}+({\cal{B}}^{(0)}(p))^{2}}.
\label{latmomt}
\eea
Results are displayed in Fig.~\ref{fig:pvsq}.

\subsection{Mass and Renormalization Functions}

The dimensionless inverse lattice bare quark propagator takes the form
\bea
S^{-1}(p)&\equiv&{ia{q\slh}A(p)+B(p)}\nm\\
&=&{[Z^{\rm{L}}(p)]^{-1}}\left({ia{q\slh}+M^{\rm{L}}(p)}\right),
\eea
where $M^{\rm{L}}(p)=B(p)/A(p)$ is the lattice quark mass function and $Z^{\rm{L}}(p)=1/A(p)$ the
lattice renormalization function. The functions $\aofp$ and $\bofp$ may be written as:
\bea
\label{ap}
\aofp&=&\frac{\caofp}{{\cal{D}}(p)},\hspace{0.3cm}{\rm{and}}\hspace{0.3cm}\bofp=\frac{\cbofp}{{\cal{D}}(p)},
\eea
where ${\cal{D}}(p)={\caofps{q^2}+\cbofps}$. Hence an equivalent definition for the quark propagator
is $S(p)\equiv{-ia{q\slh}\caofp+\cbofp}$. Extracting the functions $\caofp$ and $\cbofp$ is done via:
\bea
\label{curlap}
\caofp&=&\frac{i}{4N_caq^2}\trace[q\slh{S(p)}],\hspace{0.5cm}{\rm{and}}\hspace{0.5cm}\nm\\
\cbofp&=&\frac{1}{4N_c}\trace[S(p)].
\eea
\begin{figure}[tbh]
\resizebox*{\columnwidth}{!}{\rotatebox{90}{\includegraphics{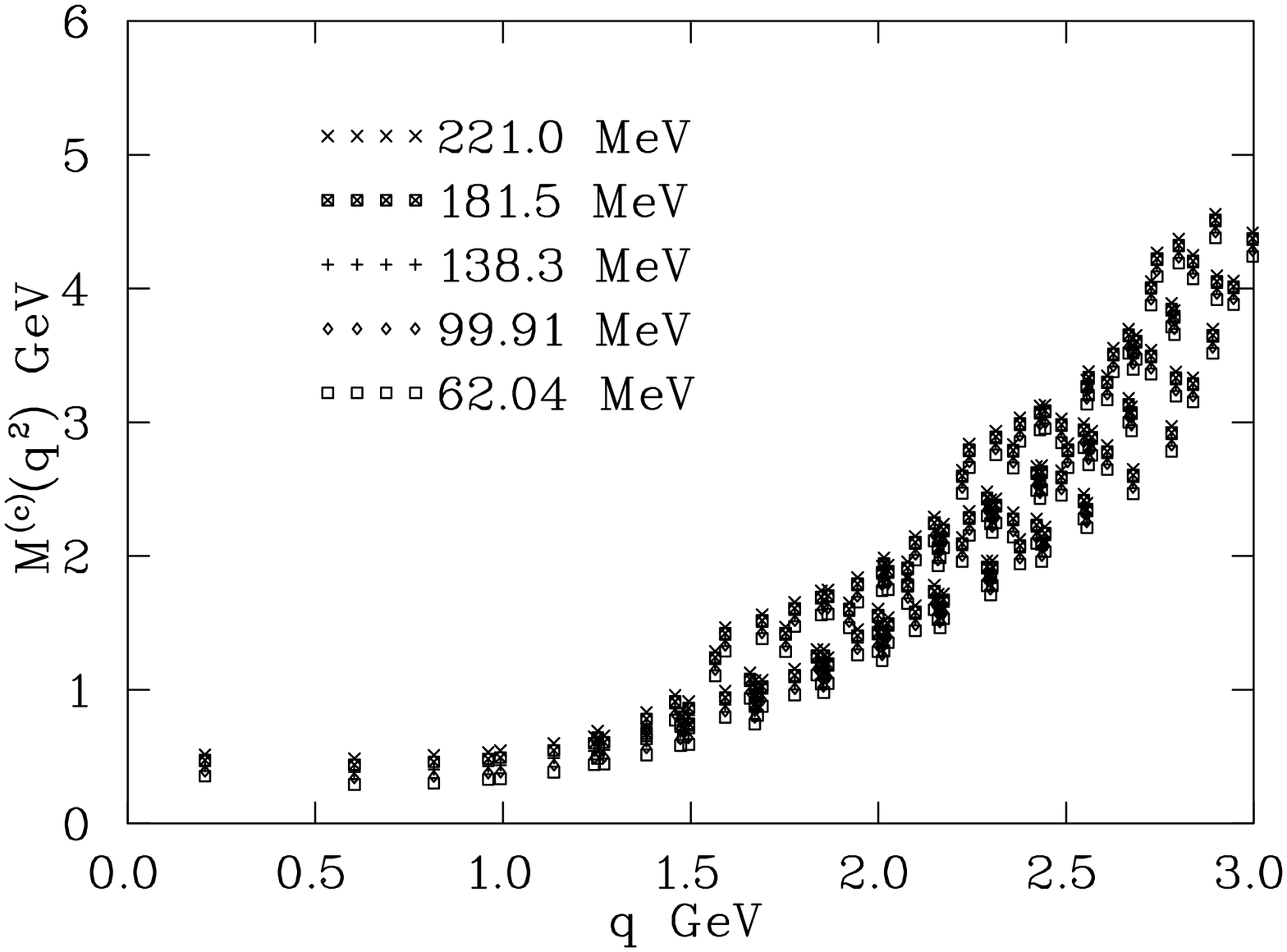}}\rotatebox{90}{\includegraphics{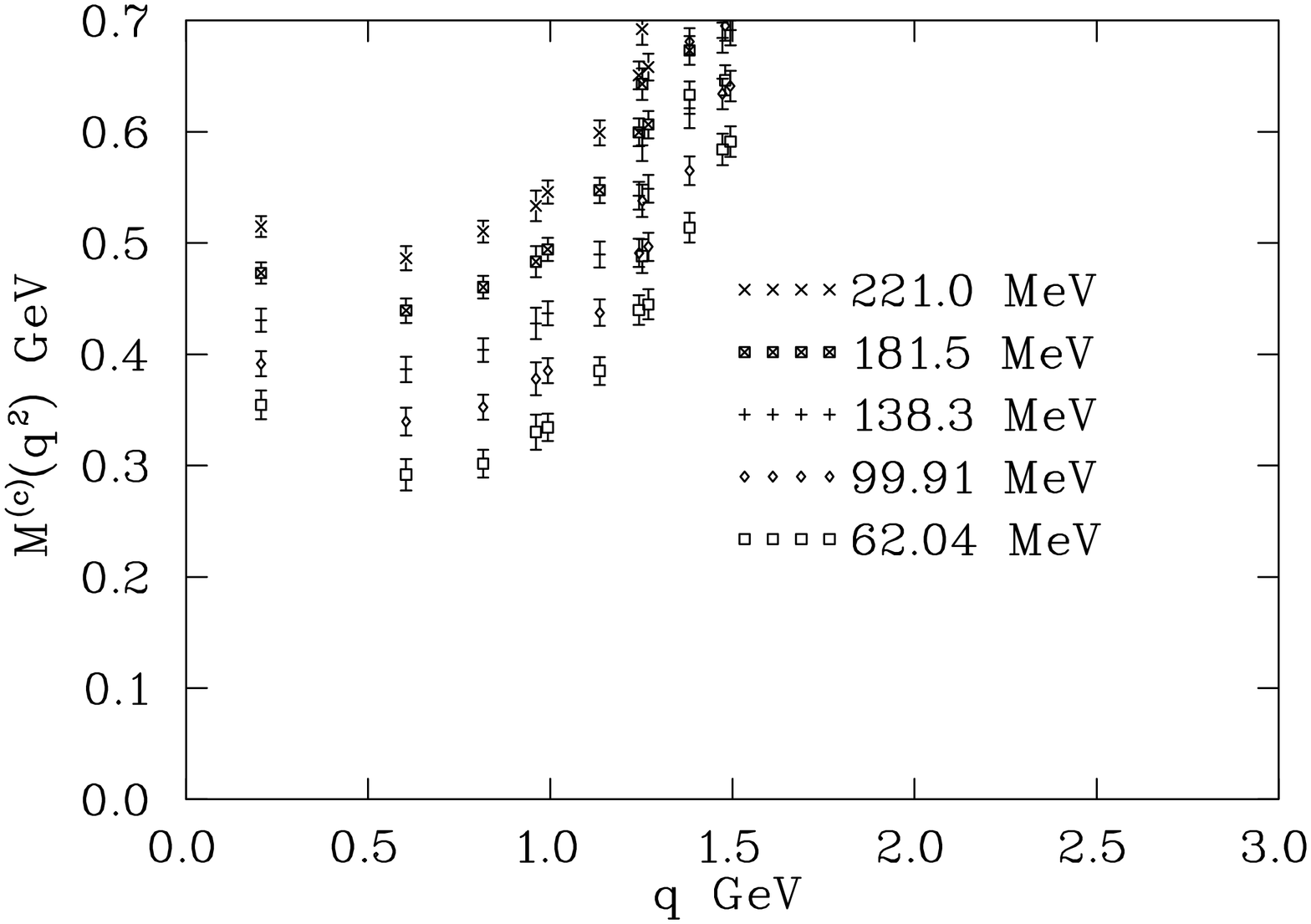}}}
\caption{\label{fig:uncorwil}The uncorrected mass function $\mofp$ for Wilson fermions.}
\end{figure}

In Fig.~\ref{fig:uncorwil} we show the uncorrected mass function for Wilson
fermions, where we can see a large divergence in the ultraviolet region.

\subsection{Tree--Level Correction}

The tree-level multiplicative correction is given by
\bea
\acofp=\frac{\aofp}{\aoofp}1\hspace{0.1cm}{\rm{,}}\hspace{0.1cm}\bcofp=\frac{\bofp}{\boofp}m_q,
\label{acbc}
\eea
where $\aoofp$ and $\boofp$ are obtained from $S^{(0)}(p)$. Then tree-level corrected mass
and renormalization functions are constructed using Eq.(\ref{acbc})
\bea
\mcofp=\left(\frac{\bcofp}{\acofp}\right)\hspace{0.1cm}{\rm{,}}\hspace{0.1cm}\zcofp=\frac{1}{\acofp}.
\eea

\section{FERMIONS ON THE LATTICE}

\subsection{Wilson Fermions}

The Wilson fermion~\cite{wilson} action on the lattice is defined as,
\bea
S_{W}[U,\overline{\psi},\psi]=\sum_{xy}\overline{\psi}(x)D_{W}(x,y)\psi(y),\nm
\label{wilfermact}
\eea
with $D_{W}(x,y)$ the usual Wilson fermion operator.
The lattice bare mass is related to the hopping parameter via 
$\kappa={1}/{(2m_q a +8r)}$ and
$m_q\equiv({1}/{2a})({1}/{\kappa}-{1}/{\kappa_{c}})$.

\subsection{Overlap Fermions}

The overlap fermion
\cite{Neu95} 
realizes exact chiral symmetry on the lattice
\bea
D(\mu)&=&\frac{1}{2}\left[1+\mu+(1-\mu)\frac{D_{W}}{\sqrt{D^{\dagger}_{W}D_{W}}}\right]\nm
\eea
with $0{\leq}\mu\leq{1}$ describing fermions with a positive mass from 0 to $\infty$.
The hopping parameter is given by $\kappa={1}/{(-2m+8r)}$ and 
$m\equiv({1}/{2\kappa_{c}}-{1}/{2\kappa})$.
To describe a single massless Dirac fermion for $D(0)$ we must have $0\,{\leq}m\leq{2}$
at tree--level. The overlap propagator is given by
\bea
\tilde{D}^{-1}(\mu)=(1-\mu)^{-1}[{D}^{-1}(\mu)-1],
\eea
and is related to the continuum propagator by ${D}^{-1}_{c}(m_q)=\calz_{\psi}\tilde{D}^{-1}(\mu)$, and similarly
for the quark mass $m_q=\calz^{-1}_{m}\mu$.

At tree-level \cite{Edw99} it is found that
$\calz_{\psi}^{(0)}=\calz_{m}^{-1}=2m$, and the free propagator becomes
\bea
\left[{D}^{(0)}_{c}(m_q)\right]^{-1}=\left[\calz_{\psi}^{(0)}\tilde{D}^{(0)}(\mu)\right]^{-1},\nm
\eea
and when the interactions are turned on we have
\bea
\left[{D}_{c}(m_q)\right]^{-1}=\left[\calz_{\psi}\tilde{D}(\mu)\right]^{-1},\nm
\eea
where $\calz_{\psi}=\calz_{m}^{-1}$. 

\section{NUMERICAL RESULTS}

\begin{figure}[tb]
\resizebox*{\columnwidth}{!}{\rotatebox{90}{\includegraphics{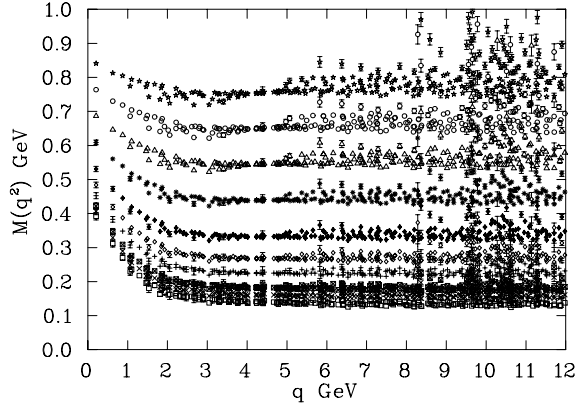}}}
\caption{\label{fig:ovrm}The mass function $\mofp$ for overlap fermions (full data) for the ten different masses.}
\end{figure}

\begin{figure}[tbh]
\resizebox*{\columnwidth}{!}{\rotatebox{90}{\includegraphics{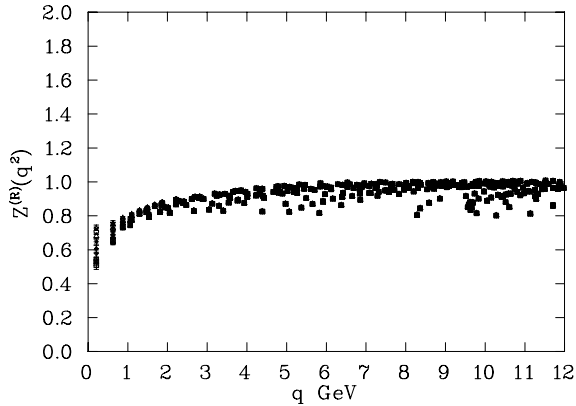}}}
\caption{\label{fig:ovrz}The renormalization function for overlap fermions (full data) for the ten different masses. The function is renormalized at $\zeta=8.2$ GeV.}
\end{figure}

\begin{figure}[tbh]
\resizebox*{\columnwidth}{!}{\rotatebox{90}{\includegraphics{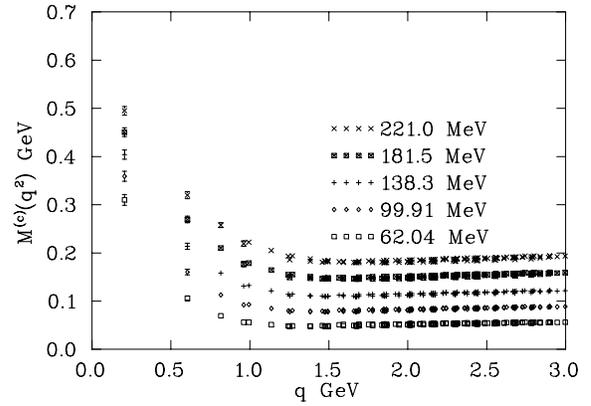}}}
\caption{\label{fig:extwilm}The tree--level corrected mass function $\mcofp$ for Wilson fermions (half cut data).}
\end{figure}

\begin{figure}[tbh]
\resizebox*{\columnwidth}{!}{\rotatebox{90}{\includegraphics{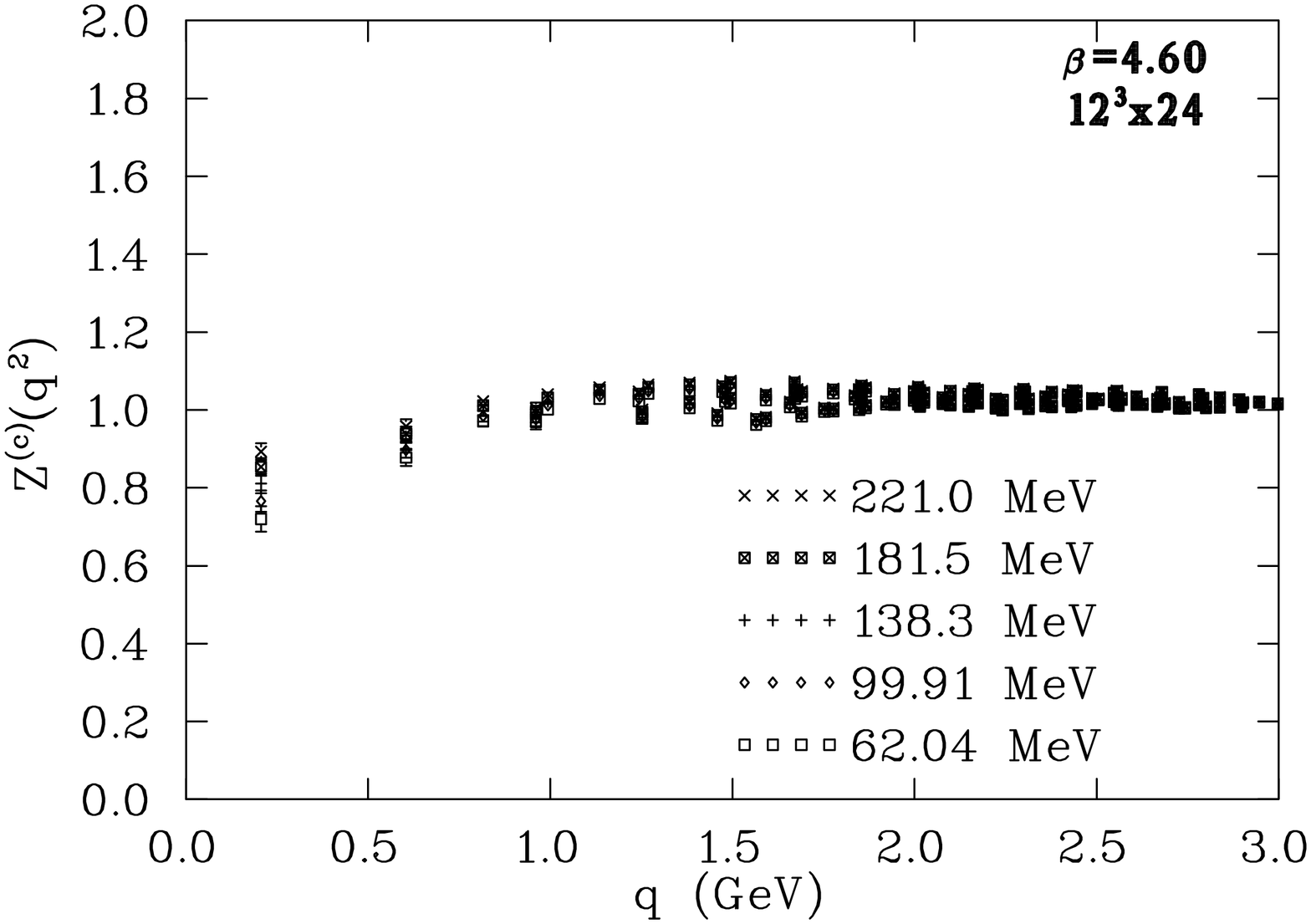}}}
\caption{\label{fig:extwilz}The tree--level corrected renormalization function for Wilson fermions (half cut data).}
\end{figure}

Using fifty improved gauge field configurations, on a $12^3\times{24}$ lattice with $a=0.125(1)$ fm, the
overlap quark propagator is calculated for ten different masses,
$m_{q}=\calz_{\psi}\mu=\{126,147,168,210,252,315,420,524,629,734\}$ MeV. Results are illustrated
in Figs.~\ref{fig:ovrm} and~\ref{fig:ovrz}.
For Wilson fermions, using a hundred configurations, five
masses are considered, namely $m_{q}=\{221,181.5,138.3,99.91,62.04\}$ MeV. Results are displayed in
Figs.~\ref{fig:extwilm} and~\ref{fig:extwilz}.

\begin{figure}[tbh]
\resizebox*{\columnwidth}{!}{\rotatebox{90}{\includegraphics{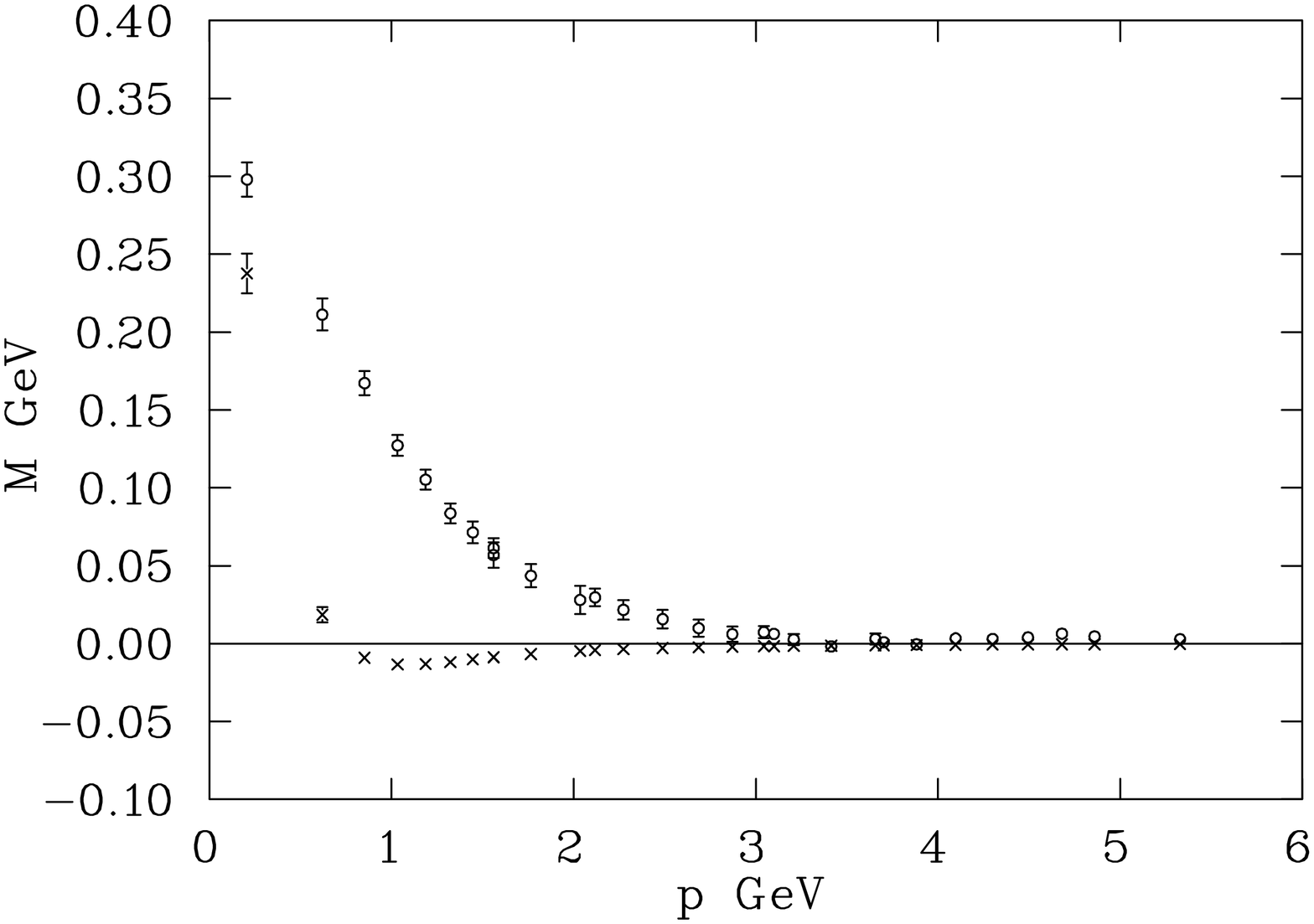}}}
\caption{\label{fig:extwilovrm}Comparison of the linearly extrapolated mass function $M$ for both Wilson fermions ($\times$) and overlap fermions ($\square$), (cylinder cut data).}
\end{figure}

A linear extrapolation is used to compare the two actions. In Fig.~\ref{fig:extwilovrm}
we show the linearly extrapolated quark mass function for both actions plotted versus the discrete
momentum $p$. When plotting $M$ and $Z$ versus the lattice momentum, $q$, the points are pushed away
from the origin in the case of overlap fermions as opposed to Wilson fermions where the points are pulled
towards the origin. The overlap action (plotted as $\square$) produces in the deep infrared $M(0)=297(11)$ MeV.
Comparing with the Wilson fermion action ($\times$) we see a clear superiority of the overlap action over the
Wilson action. Wilson fermions show a significant dip between 0.8 GeV and 2.0 GeV.

\begin{figure}[tbh]
\resizebox*{\columnwidth}{!}{\rotatebox{90}{\includegraphics{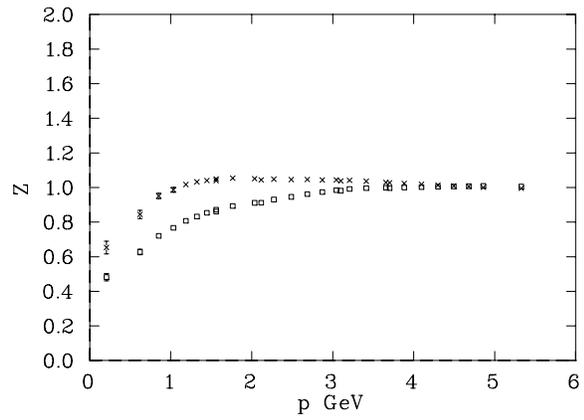}}}
\caption{\label{fig:extwilovrz}Comparison of the linearly extrapolated renormalization function for Wilson fermions ($\times$) and overlap fermions ($\square$), (cylinder cut data).}
\end{figure}

The linearly extrapolated renormalization function is shown in Fig.~\ref{fig:extwilovrz}.
In the deep infrared for the overlap action we have $Z(0)=0.48(2)$ with the renormalization point located
at $\zeta=3.9$ GeV.

\section{SUMMARY}

Tree-level correction represents a crucial step and a powerful tool for correcting
the Wilson action results. The division method produces a smooth quark mass function throughout
the momentum spectrum. We are able to gain some insights into the analytic structure of the overlap quark
propagator: $\boofp=\calz_{\psi}^{(0)}\mu$ and $\aoofp=1$. The overlap produces very good results
for $M(p)$ and $Z(p)$. No tree-level correction is required for overlap fermions.

\end{document}